\input harvmac
\def\journal#1&#2(#3){\unskip, \sl #1\ \bf #2 \rm(19#3) }
\def\andjournal#1&#2(#3){\sl #1~\bf #2 \rm (19#3) }

\def\ie{{\it i.e.}}
\def\eg{{\it e.g.}}

\def\frac#1#2{{#1\over#2}}

\def\inbar{\,\vrule height1.5ex width.4pt depth0pt}
\def\IC{\relax\hbox{$\inbar\kern-.3em{\rm C}$}}
\def\IR{\relax{\rm I\kern-.18em R}}
\def\IP{\relax{\rm I\kern-.18em P}}
\def\IZ{\relax{\rm I\kern-.18em Z}}

%
%
\def\np#1#2#3{Nucl. Phys. {\bf B#1} (#2) #3}
\def\pl#1#2#3{Phys. Lett. {\bf #1B} (#2) #3}
\def\plb#1#2#3{Phys. Lett. {\bf #1B} (#2) #3}

\def\prd#1#2#3{Phys. Rev. {\bf D#1} (#2) #3}

\catcode`\@=11
\def\slash#1{\mathord{\mathpalette\c@ncel{#1}}}
\overfullrule=0pt

\def\MM{{\cal M}}

\def\underrel#1\over#2{\mathrel{\mathop{\kern\z@#1}\limits_{#2}}}

\catcode`\@=12


%

\def\exp{{\rm exp}}


\rightline{NSF-ITP-99-104}
\rightline{RI-8-99, EFI-99-41}
\Title{
\rightline{hep-th/9909110}}
{\vbox{\centerline{Little String Theory in a Double Scaling Limit}}}
\medskip
\centerline{\it Amit Giveon${}^{1,2}$ and David Kutasov${}^{1,3}$}
\bigskip
\centerline{${}^1$Institute for Theoretical Physics}
\centerline{University of California, Santa Barbara, CA 93106, USA}
\smallskip
\centerline{${}^2$Racah Institute of Physics, The Hebrew University}
\centerline{Jerusalem 91904, Israel}
\centerline{giveon@vms.huji.ac.il}
\smallskip
\centerline{${}^3$Department of Physics, University of Chicago}
\centerline{5640 S. Ellis Av., Chicago, IL 60637, USA }
\centerline{kutasov@theory.uchicago.edu}

\bigskip\bigskip\bigskip
\noindent
A double scaling limit can be defined in
string theory on a Calabi-Yau (CY) manifold by approaching
a point in moduli space where the CY space develops an isolated
singularity and at the same time taking the string coupling to zero,
while keeping a particular combination of the two parameters
fixed. This leads to a decoupled theory without gravity which
has a weak coupling expansion, and can be studied using
a holographically dual non-critical superstring description.
The usual ``Little String Theory'' corresponds to the strong
coupling limit of this theory. We use holography to compute two
and three point functions in weakly coupled double scaled
little string theory, and study the spectrum of the theory
in various dimensions. We find a discrete
spectrum of masses which exhibits Hagedorn growth.

\vfill

\Date{9/99}

\newsec{The double scaling limit}

\lref\natistr{N. Seiberg, hep-th/9705221, \pl{408}{1997}{98}.}

In the last few years it became clear that consistency of
string theory implies the existence of new non-local theories
without gravity in six or less dimensions. One way of defining
these ``little string theories'' (LST's) is to consider string
dynamics in vacua which contain $NS5$-branes, in the decoupling
limit $g_s\to 0$ \natistr. In this limit, fluctuations in the bulk
of spacetime decouple, and one is left with the dynamics of modes
which live on the fivebranes. For example, taking this limit in
type IIA or IIB string theory in flat spacetime with $N$ fivebranes
leads to maximally supersymmetric six dimensional theories (sixteen
supercharges) with $(2,0)$ or $(1,1)$ supersymmetry, respectively.

LST's do not contain gravity but have other features that are
reminiscent of critical string theories. In particular, they have
a Hagedorn density of states, and upon compactification on tori
they exhibit T-duality. There are some apparent differences as well:
\item{(1)} Unlike critical strings, for which it is believed that
one can only study on-shell physics (\eg\ the S-matrix), LST's are
expected to have well defined off-shell Green's functions.
\item{(2)} Critical string theory in flat non-compact spacetime
has a fundamental scale, the string scale $l_s$, and a dimensionless
coupling, $g_s$. For small $g_s$, the theory is weakly coupled. LST's,
as defined above, have a scale $l_s$, and are further labeled
by the number of fivebranes $N$. They appear to be inherently
strongly coupled. One might be tempted to associate $1/N$
with a coupling in LST, but there is no known sense in which the theory
becomes weakly coupled at large $N$ (it does become weakly coupled at
large $N$ for sufficiently low energies). Nevertheless, we will see
that one can in fact define LST's which are arbitrarily weakly
coupled, so in this respect LST is actually similar to critical
string theory.

\noindent
Another definition of LST, which will be more useful for our
purposes, involves the study of string vacua of the form
$\IR^{d-1,1}\times \MM$, where $\MM$ is a CY manifold with an
isolated singularity, in the decoupling limit $g_s\to 0$.
Since in this limit all non-trivial physics
is localized near the singularity, to study the resulting
LST one can replace the CY
manifold $\MM$ by its form near the singular point.

\lref\OV{H. Ooguri and C. Vafa, hep-th/9511164, \np{463}{1996}{55}.}%
\lref\Kutasov{D. Kutasov, hep-th/9512145, \plb{383}{1996}{48}.}%

In the maximally supersymmetric, six dimensional
case of \natistr, $\MM=K3$ and the possible singularities
have an ADE classification. The theory on $N$ $NS5$-branes
mentioned above corresponds to an $A_{N-1}$ singularity,
near which the $K3$ manifold looks like an ALE
space -- the singular, non-compact hypersurface
\eqn\anmo{z_1^N+z_2^2+z_3^2=0}
in $\IC^3$.
The two descriptions of six dimensional LST given above
are related by T-duality \refs{\OV,\Kutasov}. The
theory on $N$ type IIA $NS5$-branes is equivalent to
the IIB theory on $\IR^{5,1}$ times the ALE space \anmo,
and vice versa.

An interesting generalization of the above procedure is
the following. Let $t$ be one of the moduli of $K3$ CFT,
and $t=t_c$ be the value of $t$ for which the $K3$ surface
develops an isolated singularity. The standard LST limit
is $t\to t_c$, followed by $g_s\to 0$. Instead, one can
consider a double scaling limit, \ie\ take $t\to t_c$
and $g_s\to 0$ at the same time, keeping some combination,
$(t-t_c)^\gamma/g_s$,  fixed.

\lref\gkp{A. Giveon, D. Kutasov
and O. Pelc, hep-th/9907178.}%
\lref\ginsmoore{For a review, see P. Ginsparg and G. Moore,
hep-th/9304011.}%
\lref\gv{S. Mukhi and C. Vafa, hep-th/9301083, Nucl. Phys. 
{\bf B407} (1993) 667; D. Ghoshal and C. Vafa, hep-th/9506122,
Nucl. Phys. {\bf B453} (1995) 121.}%

To describe this limit more precisely it is convenient to
blow up the vicinity of the singularity and replace the
$K3$ surface by the ALE space \anmo\ describing the region
near the singularity (at $z_a=0$). One way of smoothing the
singularity is to replace \anmo\ by
\eqn\ab{z_1^N+z_2^2+z_3^2=\mu~,}
where $\mu$ can be thought of as one of the moduli of the
underlying $K3$ CFT ($\mu\sim t-t_c$ in the previous notation).
As we will see later, string theory on $\IR^{5,1}$ times
the surface \ab\ depends on $\mu$, $g_s$ only through the scaling
parameter $x\equiv \mu^{1\over N}/g_s$. The double scaling limit
corresponds to $\mu\to 0$, $g_s\to 0$ with $x$ held fixed.
Note that:
\item{(1)} The string coupling expansion corresponds in this
case to an expansion in powers of $1/x$.
\item{(2)} The original definition of LST corresponds to
$x=0$ ($\mu\to 0$ first, then $g_s\to 0$), and hence gives
rise to a strongly coupled theory.
\item{(3)} While in string theory on $K3$ one
actually has to take the double scaling limit to decouple
the LST from gravity and bulk string theory, replacing
the $K3$ surface by \ab\ automatically implements the limit. In
particular, $\mu$ does not have to be small in \ab; in fact, the LST
becomes weakly coupled at {\it large} $\mu$.
\item{(4)} As emphasized in \gkp, the double scaling limit defined
above is qualitatively very similar to the one used in the study
of two dimensional critical string theory \ginsmoore, and one can borrow
much of the intuition developed there.
\item{(5)} String theory on resolved ALE spaces was considered
in some very interesting pre-holography papers \refs{\gv,\OV}.
We will comment later on the relation of these papers to our work.

\noindent
The deformation \ab\ has a natural interpretation in terms of
the theory of $N$ $NS5$-branes\foot{We thank O. Aharony and T.
Banks for discussions of this issue.}. Rewriting \ab\ as
\eqn\deffive{\prod_{j=1}^N(z_1-r_0e^{2\pi ij\over N})+z_2^2+z_3^2=0;
\;\;\mu=(-r_0)^N}
and performing the T-duality of \refs{\OV,\Kutasov}, we find
a system of $N$ fivebranes distributed uniformly on a circle
of radius $r_0$, which is related to $\mu$ as in \deffive.
In type IIB string theory, the low energy theory on $N$ $NS5$-branes
is $SU(N)$ SYM theory with sixteen supercharges, and \deffive\
corresponds to a point in the moduli space of vacua of the gauge
theory, in which a complex scalar field $\Phi$ in the adjoint
representation of the gauge group has a v.e.v.
\eqn\pointmod{\langle\Phi\rangle=C{\rm diag} (e^{2\pi i\over N},
e^{4\pi i\over N},\cdots, e^{2\pi iN\over N})~.}
The double scaling limit $\mu, g_s\to 0$ with $x=\mu^{1\over N}/g_s$
fixed has a nice interpretation in this language. At the point
\pointmod\ in moduli space, the $SU(N)$ gauge symmetry is broken
to $U(1)^{N-1}$, and the mass of the off-diagonal gluons (``W-bosons'')
is $m_W\sim r_0/g_sl_s\sim x/l_s$ (since W-bosons correspond to $D1$-branes
stretched between the fivebranes).

Thus, the double scaling limit is simply the decoupling limit of
\natistr\ with fixed W-boson masses. The scaling parameter $x$
is the scale of these masses (in string units). The double scaled
theory is weakly coupled for energies well below $m_W$. For
$E>>m_W$ one expects to recover the strongly coupled LST at
the origin \natistr. 
A similar story (with slightly different details) can be told for
type IIA fivebranes.

The above discussion can be generalized to higher dimensional
Calabi-Yau spaces. One again studies the limit where
the CY manifold develops an isolated singularity and blows up
the region near the singularity to define the double scaling
limit. There are many possible singularities that
can appear on CY manifolds. We will restrict
our attention to quasi-homogenous
hypersurface singularities, in which the vicinity of the
singularity is (as in \anmo) a hypersurface $F(z_1,\cdots,
z_{n+1})=0$ in $\IC^{n+1}$,
where $F$ is a quasi-homogenous polynomial with weight one under
$z_a\to\lambda^{r_a}z_a$, \ie\
\eqn\homog{F(\lambda^{r_a}z_a)=\lambda F(z_a)~, \;\;\lambda\in \IC~,}
for some set of positive weights $r_a$. The fact that the singularity
is isolated implies that $F$ is
transverse, \ie\ the only point at which all derivatives
$\partial_{z_a} F$ vanish is the origin, $z_a=0$.
The double scaling limit gives rise in this case to a
$d=10-2n$ dimensional non-gravitational theory. It involves string theory
on $\IR^{d-1,1}$ times the non-compact manifold
\eqn\noncomp{F(z_1,\cdots, z_{n+1})=\mu~.}
We will see below that the combination
of $\mu$ and $g_s$ on which physics in this background
depends is the scaling variable
\eqn\scvar{x\equiv {\mu^{r_\Omega}\over g_s}~,}
where
\eqn\rom{r_\Omega\equiv\sum_{a=1}^{n+1} r_a-1~.}
We will restrict to $r_\Omega>0$. The resulting
LST is weakly coupled for large $x$ and we will
study it in this limit.

\newsec{The holographic description}

\lref\adscft{For a review, see O. Aharony, S. Gubser, J. Maldacena,
H. Ooguri and Y. Oz, hep-th/9905111.}%
\lref\abks{O. Aharony, M. Berkooz, D. Kutasov and N. Seiberg,
hep-th/9808149, JHEP {\bf 9810} (1998) 004.}%
\lref\imsy{N. Itzhaki, J. Maldacena, J. Sonnenschein and S.
Yankielowicz, hep-th/9802042, \prd{58}{1998}{046004}.}%
\lref\bst{H. J. Boonstra, K. Skenderis and P. K. Townsend,
hep-th/9807137, JHEP {\bf 9901} (1999) 003.}%
\lref\dg{D. Diaconescu and J. Gomis, hep-th/9810132,
\np{548}{1999}{258}.}%
\lref\kll{D. Kutasov, F. Larsen and R. Leigh, hep-th/9812027,
\np{550}{1999}{183}.}%
\lref\minsei{S. Minwalla and N. Seiberg,  hep-th/9904142,
JHEP {\bf 9906} (1999) 007.}%
\lref\gk{M. Gremm and A. Kapustin, hep-th/9907210.}%

The construction described in the previous section
is useful for establishing the existence of LST's, but
it does not provide efficient techniques for studying
them in detail. To go further, we will use a holographic description
of these theories. This description, which is a generalization of
the AdS/CFT correspondence \adscft, was proposed in \abks\ (see also
\refs{\imsy,\bst,\dg,\kll,\minsei,\gk}) and was generalized to the case of
general hypersurface singularities \noncomp\ in \gkp. We will next
briefly summarize it.

The claim is that holography relates the decoupled dynamics at the
singular point $z_a=0$ in string theory on $\IR^{d-1,1}$
times the space $F(z_a)=0$ \homog\ (``the boundary theory'') to
a ``bulk theory,'' string theory on
\eqn\rdd{\IR^{d-1,1}\times\IR_\phi\times S^1\times LG(W=F)~,}
where $\IR_\phi$ is the real line, labeled by $\phi$, along which
the dilaton varies linearly,
\eqn\lindil{\Phi=-{Q\over2}\phi~,}
and $LG(W=F)$ is a Landau-Ginzburg $N=2$ SCFT of $n+1$ chiral
superfields $z_a$ with superpotential $W(z_a)=F(z_a)$ \homog.
It is easy to check \gkp\ that consistency of string propagation on
\rdd\ requires a relation between $r_\Omega$ \rom\ and $Q$ \lindil:
\eqn\Qrom{Q^2=2r_\Omega~.}
We will label the $S^1$ in \rdd\ by $Y$; its radius is determined
by the GSO projection to be $Q$. The GSO projection further
acts as an orbifold by a shift along the circle, $Y\to Y+\pi Q$,
and (roughly) as fermion number on the rest of the theory (see
\gkp\ for the details).

\lref\chs{C. Callan, J. Harvey and A. Strominger, hep-th/9112030,
in Trieste 1991, Proceedings, String Theory and Quantum Gravity 1991,
208.}

For the six dimensional, maximally supersymmetric case \ab,
the superpotential is $W=z_1^N+z_2^2+z_3^2$, and the LG model
is an $N=2$ minimal model, which can be thought of as the
coset SCFT $SU(2)_N\over U(1)$. The GSO projection acts as a
$Z_N$ orbifold on ${SU(2)_N\over U(1)}\times S^1$, turning
it into $SU(2)_N$. The backgound \rdd\
thus becomes $\IR^{5,1}\times \IR_\phi\times SU(2)$, recovering
the results of \chs.

\lref\ks{D. Kutasov and N. Seiberg, \pl{251}{1990}{67}; for a review,
see D. Kutasov, hep-th/9110041.}

The theory described by \rdd\ is singular. The string coupling
$g_s\sim \exp(-Q\phi/2)$ blows up as $\phi\to-\infty$; the
theory runs to strong coupling. This is not surprising in
view of the earlier discussion, since \rdd\ is dual to the
singular background \noncomp\ with $\mu=0$, or \scvar\
$x=0$, which as we mentioned is strongly coupled.
One would like to turn on $\mu$ in \noncomp\ and study the
dual theory for large $x$, where it is weakly coupled.
Turning on $\mu$ in \noncomp\ corresponds \refs{\ks,\gkp}
in the dual theory \rdd\ to adding to the worldsheet
Lagrangian the superpotential term
\eqn\ntwosup{\delta\CL=\mu\int d^2\theta e^{-{1\over Q}(\phi+iY)}+c.c.~.}
Here $\phi$, $Y$ are worldsheet superfields whose bosonic
components are the scalar fields $\phi$, $Y$ discussed above.
A few comments about the interaction \ntwosup\ are in order:
\item{(1)} This is the $N=2$ Liouville interaction. It preserves
worldsheet $N=2$ superconformal symmetry and hence spacetime
supersymmetry. At the same time it prevents the dilaton from
running to strong coupling.
\item{(2)} Even before we add \ntwosup\ to the Lagrangian,
the fact that this operator is in the spectrum of the theory
implies that the radius of $Y$ is an integer multiple of $Q$.
An analysis of the rest of the spectrum shows \gkp\ that this
radius is equal to $Q$.
\item{(3)} We can now derive \scvar. Standard scaling arguments
(see \eg\ \ginsmoore) imply that the perturbative expansion
in the vacuum \rdd, \lindil, \ntwosup\ is a series in $g_s/\mu^{Q^2\over2}$.
Using \Qrom, this leads to \scvar.

\noindent
It is interesting to compare our discussion to the work of \refs{\gv,\OV}.
In these papers it was pointed out that the sigma-model whose target space
is  the non-compact manifold \noncomp\ can be alternatively described
by a Landau-Ginzburg model with superpotential
\eqn\supgv{W=-\mu z_0^{-k}+F(z_1,\cdots, z_{n+1})~,}
where $z_0$ is an additional chiral superfield, and
\eqn\kkk{k\equiv 1/r_\Omega~.}
The description \supgv\ is natural since the equation $W(z_0,z_a)=0$ in
$n+1$ dimensional weighted projective space satisfies the CY condition
with $\hat c=n$, and if $z_0\not=0$ we can set it to one and recover
the space \noncomp. It was also argued in \refs{\gv,\OV} that the patch
with $z_0=0$ should not appear since in the LG phase the superpotential
\supgv\ pushes $z_0$ to large values.

\lref\witcig{E. Witten, Phys. Rev. {\bf D44} (1991) 314.}%

The first term in the superpotential \supgv\ appears to
be ill defined: the corresponding potential does not have
a minimum at a finite value of $z_0$, and in general $k$
is non-integer, which makes $W$ non single valued. In
\refs{\gv,\OV} it was proposed to interpret it as an $SL(2)/U(1)$
coset SCFT at level $k$, and arguments were presented to support
this interpretation. Geometrically, this coset corresponds to
a semi-infinite cigar \witcig, with the string coupling going to zero
far from the tip and approaching an arbitrary
finite value at the tip ($g_s\sim \mu^{-1/k}$).

Comparing \rdd, \ntwosup\ with \supgv\ we see that our description
of \noncomp\ is very similar, but appears to be different from
that of \refs{\gv,\OV}. For $\mu=0$ the two are identical: both
contain a LG model with superpotential $W=F$, and an infinite cylinder
$\IR_\phi\times S^1$ with a linear dilaton along $\IR_\phi$. When
$\mu$ is turned on, the two descriptions naively disagree. In our
case, the strong coupling singularity at $\phi\to-\infty$ is cut off
by a superpotential \ntwosup, whereas in \refs{\gv,\OV} it is
eliminated by changing the topology of the cylinder
to that of the semi-infinite cigar.

Since both descriptions are quite well motivated,
we would like to propose that they are in fact
equivalent, \ie\ that $N=2$ Liouville
(with a cosmological constant $\mu$) and $SL(2)/U(1)$
(with $e^{\Phi}\sim \mu^{-{1\over k}}$ at the tip of the cigar)
are isomorphic SCFT's. They
are related by strong-weak coupling duality on the worldsheet
(possibly, a kind of T-duality).
We have not proven this statement but
would like to offer the following comments in its support:
\item{(1)} The matching of the central charges of the two theories
relates the slope of the linear dilaton $Q$ \lindil\ to the level
$k$ of $SL(2)/U(1)$. The relation, which follows from \Qrom, \kkk,
is $Q^2=2/k$. Thus, if the two are equivalent, the relation between
them is indeed a strong-weak coupling duality. $N=2$ Liouville
is weakly coupled for large Liouville central charge, \ie\ large
$Q$, while $SL(2)/U(1)$ is weakly coupled for large $k$ (when
the curvature of the cigar goes to zero everywhere).
\item{(2)} Recall that the radius of the $S^1$ (labeled by $Y$)
in $N=2$ Liouville is $Q$. The asymptotic radius of the Euclidean
cigar is $\sqrt{2k}$ \witcig. The two are indeed inverses of each
other, $R_Y=2/R_{\rm cigar}$, as appropriate for T-dual theories.
\item{(3)} In $N=2$ Liouville, the cosmological term breaks
translation invariance in $Y$. In contrast, in $SL(2)/U(1)$
translation invariance around the cigar is not broken, but winding
number is not a good symmetry, since winding modes around the cigar
can move to the tip and contract to a point. This again is consistent
with an interpretation of the relation between the two theories as
T-duality, which exchanges winding and momentum modes. In both
theories, as $\mu\to 0$ the respective symmetries are restored.
\item{(4)} A closely related duality to the one advocated here
appeared in unpublished work by V. A. Fateev, A. B. Zamolodchikov and
Al. B. Zamolodchikov. It relates
the bosonic $SL(2)/U(1)$ CFT to a bosonic analog of our $\IR_\phi
\times S^1$ with the superpotential \ntwosup\ replaced by a
bosonic potential of the form $e^{\beta\phi}\cos(\gamma Y)$.

\lref\gpr{For a review, see A. Giveon, M. Porrati and E.
Rabinovici, hep-th/9401139, Phys. Rept. {\bf 244} (1994) 77.}%
\lref\ghm{R. Gregory, J. Harvey and G. Moore, hep-th/9708086,
Adv. Theor. Math. Phys. {\bf 1} (1997) 283.}%

\noindent
Finally, we should remark on the possible relation of the duality proposed
above to the standard T-duality with respect to the Abelian isometry
of the cigar CFT \gpr. The latter takes the cigar into a trumpet,
which seems different from the $N=2$ Liouville background at small $\phi$.
However, it might be that non-perturbative worldsheet effects modify the
trumpet background.
The Abelian T-duality of an ALE space, mentioned previously,
is an example where the naive dual background is expected to receive
worldsheet instanton corrections which modify it to the (near-horizon
geometry of the) $NS5$-brane background (more generally, the Abelian dual 
of a Kaluza-Klein monopole is expected to receive worldsheet instanton
corrections which modify it to an H-monopole \ghm).

\newsec{Correlation functions and spectrum}

In this section we will use the holographic description
of \refs{\abks,\gkp} to compute correlation functions
in weakly coupled double scaled little string theory. Recall that, in
general, holography relates observables in the non-gravitational
``boundary theory'' to non-normalizable on-shell states in the
``bulk theory.'' A large class of such observables corresponds
to non-normalizable vertex operators (\ie\ fundamental string
states) in the bulk theory, and we will focus on those here.

To get started, we need to decide which of the two T-dual
descriptions of the bulk theory to use, the $SL(2)/U(1)$
or $N=2$ Liouville one. As explained in section 2,
the two are supposed to be equivalent, but each is more
appropriate in a different region of parameter space.
For large $k$ it is better to use $SL(2)/U(1)$, while
for small $k$ the $N=2$ Liouville description provides
a better qualitative guide to the physics. The transition
between the two descriptions occurs at $k=1$ (corresponding 
to the self dual radius $R_Y=R_{\rm cigar}=\sqrt2$).

In our context, $k$ lies in the range\foot{The lower bound corresponds
to pure $N=2$ Liouville or superconformal $SL(2)/U(1)$,
where $c=3+6/k=15\Rightarrow k=1/2$.} $1/2\le k<\infty$.
For example, in the maximally supersymmetric six dimensional
case we have $k=N$, the number of fivebranes, and it can become
arbitrarily large, but is bounded from below by two.
Therefore, it is natural to use the $SL(2)/U(1)$ variables
to describe the observables and compute correlation functions.
We next recall a few facts about $SL(2)/U(1)$ CFT.

\subsec{$SL(2)/U(1)$ CFT}

A convenient description of the $SL(2)/U(1)$ coset CFT
is obtained by starting with $SL(2)\times U(1)$ and
gauging the null diagonal $U(1)$ obtained from $J^3$ of
$SL(2)$ and the extra $U(1)$ (and similarly for the
other worldsheet chirality). Realizing the extra $U(1)$
in terms of a canonically normalized free scalar field $Y$,
we gauge the axial symmetry
\eqn\Jga{(J^3-i\sqrt{k\over2}\partial Y,
-\bar J^3-i\sqrt{k\over2}\bar\partial \bar Y)~,}
which leads to the cigar geometry.
Observables in the coset theory are obtained from
observables in $SL(2)\times U(1)$ CFT by imposing
invariance under \Jga\ (and identifying gauge
equivalent observables). Thus, one can use results
about correlation functions in $SL(2)$ CFT to study
the coset.

\lref\ksp{D.\ Kutasov and N.\ Seiberg, hep-th/9903219,
JHEP {\bf 9904} (1999) 008.}%

Observables in $SL(2)$ CFT\foot{More precisely, we will
be discussing the Euclidean analog of the $SL(2)$ group
manifold, $H_3^+=SL(2,\IC)/SU(2)$.} (see \eg\ \ksp\ for
a recent discussion) are obtained by studying functions
on the manifold $H_3^+$ and then applying to them the generators
of the current algebra. A convenient basis for such
functions is
\eqn\ooyy{\Phi_j(x,\bar x)={2j+1\over\pi}
\left({1 \over (\gamma-x)(\bar \gamma -\bar x)e^\phi+e^{-\phi}}
\right)^{2(j+1)},}
where $(\phi,\gamma,\bar\gamma)$ are coordinates on $H_3^+$
(the metric is ${1\over k}ds^2=d\phi^2+e^{2\phi}d\gamma d\bar\gamma$)
and $(x,\bar x$) are auxiliary variables introduced
in \ref\zamfat{A. B. Zamolodchikov and V. A. Fateev,
Sov. J. Nucl. Phys. {\bf 43} (1986) 657.}. In the quantum
CFT on $H_3^+$, the $\Phi_j$ are primaries of the full
$SL(2)\times SL(2)$ current algebra; their scaling dimensions
are
\eqn\scdims{\Delta_j=\bar\Delta_j=-{j(j+1)\over k-2}~.}
For future reference we record their behavior for large $\phi$
(see \ksp\ for a more detailed discussion),
\eqn\lrgphi{\Phi_j\simeq e^{2j\phi}\delta^2(\gamma-x)+
\cdots}
For studying the coset it is convenient to ``Fourier transform''
the fields $\Phi_j(x,\bar x)$ and define the mode operators
\eqn\modeops{\eqalign{
\Phi_{j;m,\bar m}=&\int d^2x\, x^{j+m}\bar x^{j+\bar m}
\Phi_j(x,\bar x)~,\cr
\Phi_j(x,\bar x)=&{1\over V}\sum_{m,\bar m}\Phi_{j;m,\bar m}
x^{-m-j-1}\bar x^{-\bar m-j-1}~,\cr
}}
where $V\equiv\int {d^2x\over|x|^2}$ is the volume of the boundary
of $H_3^+$.
The OPE algebra of the mode operators with the $SL(2)$ currents is
\eqn\onea{\eqalign{
J^3(z) \Phi_{j;m,\bar m}(w,\bar w)\sim &{m \Phi_{j;m,\bar m}\over z-w}~,\cr
J^\pm(z) \Phi_{j;m,\bar m}(w,\bar w)\sim &{(m\mp j)\Phi_{j;m\pm 1,\bar m}
\over z-w}~,\cr }}
and similarly for $\bar J^A(\bar z)$.

A set of observables in the coset theory is obtained
by coupling $\Phi_{j;m,\bar m}$ to $Y$,
\eqn\obscoset{V_{j;m,\bar m}\equiv \Phi_{j;m,\bar m}
e^{i\sqrt{2\over k}(mY-\bar m\bar Y)}~.}
Note that $V_{j;m,\bar m}$ is not charged under \Jga\
and hence is physical. Its scaling dimensions are
\eqn\scdimv{\Delta_{j;m,\bar m}=-{j(j+1)\over k-2}+{m^2\over k};\;\;
\bar\Delta_{j;m,\bar m}=-{j(j+1)\over k-2}+{\bar m^2\over k}~,}
where $(m,\bar m)$ run over a set of momenta and windings consistent
with the fact that the radius of $Y$ is $\sqrt{2k}$ \witcig:
\eqn\mbarm{m={1\over2}(n_1+n_2k);\;\;\bar m=-{1\over2}(n_1-n_2k)~.}
One can think of $n_1$ as the momentum around the cigar (which
is conserved) and $n_2$ as the winding (which is not).

\lref\morerefs{
J. Balog, L. O'Raifeartaigh, P. Forgacs, and A. Wipf,
Nucl. Phys. {\bf B325} (1989) 225;
L.J. Dixon, M.E. Peskin and J. Lykken,  Nucl. Phys. {\bf B325} (1989) 329;
P.M.S. Petropoulos, Phys. Lett. {\bf B236} (1990) 151;
I. Bars and D. Nemeschansky, Nucl. Phys. {\bf B348} (1991) 89;
S. Hwang, Nucl. Phys. {\bf B354} (1991) 100;
K. Gawedzki, hep-th/9110076;
I. Bars, Phys. Rev. {\bf D53} (1996) 3308, hep-th/9503205;
in {\it Future Perspectives In String Theory} (Los Angeles, 1995),
hep-th/9511187;
J.M. Evans, M.R. Gaberdiel, and M.J. Perry, hep-th/9812252.}%

Not all values of $j$ are allowed in this theory. Non-normalizable
vertex operators $V_{j;m,\bar m}$ correspond to real $j> -1/2$.
Furthermore, unitarity implies that only operators with $j<(k-2)/2$
are kept \morerefs. Unitarity also implies that all scaling
dimensions of Virasoro primaries in the CFT should be
positive (except for the identity);
in particular, \scdimv\ must satisfy $\Delta_{j;m,\bar m},
\bar\Delta_{j;m,\bar m}>0$.

At large positive $\phi$ the form of the observables \obscoset\
simplifies. Using \lrgphi, \modeops, \obscoset\ we find that
\eqn\lrgphiobs{V_{j;m,\bar m}\sim e^{2j\phi}\gamma^{j+m}
\bar\gamma^{j+\bar m}e^{i\sqrt{2\over k}(mY-\bar m\bar Y)}~.}
As explained in \ref\bersh{M. Bershadsky and D. Kutasov,
Phys. Lett. {\bf 266B} (1991) 345.}, in this region the theory
simplifies and the target space can be thought of as a cylinder,
$\IR_\phi\times S^1$, labeled by $\phi$ and $Y$. The powers
of $\gamma$, $\bar\gamma$ in \lrgphiobs\ can be dropped and
$V_{j;m,\bar m}$ go over to the standard observables in a linear
dilaton vacuum.

\lref\teschner{J. Teschner, hep-th/9712256; hep-th/9712258;
hep-th/9906215; V.A. Fateev, A.B. Zamolodchikov and Al.B. 
Zamolodchikov, unpublished.}%

We will be interested below in a supersymmetric version of the
above coset CFT. This leads to a few small modifications in the
analysis. If we denote by $k$ the total level of the $SL(2)$
current algebra on $H_3^+$, then the level of the bosonic part
of the algebra is $k+2$ (with the fermions contributing
$-2$ units). Thus, the scaling dimension of the primaries $\Phi_j$
\ooyy\ is shifted by $k\to k+2$ and the unitarity bound of
\morerefs\ becomes
\eqn\unitbound{j<{k\over2}~.}
Since the $U(1)$ current $J^3$
that we are modding out by still has level $k$, formulae like
\Jga\ remain unchanged. Thus, in the superconformal coset
$SL(2)/U(1)$ we have observables $V_{j;m,\bar m}$ given by
\obscoset\ (with $m$, $\bar m$ satisfying \mbarm) whose
scaling dimensions are
\eqn\scdimvsus{\Delta_{j;m,\bar m}={m^2-j(j+1)\over k};\;\;
\bar\Delta_{j;m,\bar m}={\bar m^2-j(j+1)\over k}~.}
Positivity of the scaling dimensions \scdimvsus\ implies in this
case the constraint
\eqn\posdim{m^2,\bar m^2>j(j+1)~.}
Two and three point functions of the $SL(2)$ primaries
\ooyy\ in $H_3^+$ CFT
were computed in \teschner. The two point function is
\eqn\twoptphi{\langle\Phi_{j_1}(x_1;z_1)\Phi_{j_2}(x_2;z_2)\rangle=
C(j_1;k)|x_1-x_2|^{-4(j_1+1)}|z_1-z_2|^{-4\Delta_{j_1}}
\delta(j_1-j_2){\Gamma(1-{2j+1\over k})\over\Gamma({2j+1\over k})}~,}
where $C(j;k)$ is a numerical factor which can be found in \teschner.
Since we will be primarily interested in the analytic structure
of the two point function, it is sufficient to note that $C(j;k)$
does not have poles or zeroes for finite $j$ and $k>1$.
We have written \twoptphi\ in a notation
suited for application to the supersymmetric case, \ie\ the
level of the bosonic $\widehat{SL(2)}$ implied in \twoptphi\
is $k+2$. The two point function \twoptphi\
does not have any singularities
for all $j$ satisfying the inequality
\eqn\imprbound{-{1\over2}<j<{k-1\over2}~.}
As $j$ approaches the lower bound in \imprbound, the
two point function goes to zero, while when it approaches
the upper bound it diverges.

Clearly, for physical applications
(\eg\ for string theory on $AdS_3$)
both types of singularities are undesirable.
For example, outside of the range \imprbound\
states in the spacetime CFT of string
theory on $AdS_3$ have negative norm. Fortunately,
the lower bound in \imprbound\ is precisely the
condition for non-normalizability of the vertex
operators discussed above, while the upper bound
almost coincides with the unitarity constraint
\unitbound. In fact, \imprbound\ is slightly
stronger. Note that
this is not in contradiction with \morerefs. In these
papers it is assumed that the current algebra primary
state has positive norm, and one asks whether there are
any negative norm states in the current algebra block
obtained by acting on the primary with creation operators
$J^A_{-n}$ (after modding out $U(1)$).
The two point function \twoptphi\ instead
probes the question whether the norm of the primaries changes
sign as one varies $j$. Thus, below we will impose \imprbound\
on the observables.

Using \twoptphi\ one can now compute the two point functions
of observables in the coset theory \obscoset. The first step
is to transform from the ``local fields'' $\Phi_j(x,\bar x)$
\ooyy\ to the modes \modeops. To simplify the formulae
we will set $m=\bar m$, \ie\ restrict to pure winding modes
around the cigar \mbarm. One finds
\eqn\twoptmode{\eqalign{
&\langle\Phi_{j_1;m,m}\Phi_{j_2;-m,-m}\rangle=\cr
&\pi\delta(j_1-j_2)C(j_1;k)
{\Gamma(1-{2j+1\over k})\Gamma(1+j+m)\Gamma(j-m+1)\Gamma(-2j-1)
\over
\Gamma({2j+1\over k})\Gamma(-j-m)\Gamma(m-j)\Gamma(2j+2)}\cr}}
(here and below we supress the $z$ dependence).

To compute the two point function of coset observables $V_{j;m,m}$
\obscoset\ we multiply \twoptmode\ by the two point function
of the exponential in $Y$. This does not change the result;
hence we find
\eqn\twoptcoset{\eqalign{
&\langle V_{j_1;m,m}V_{j_2;-m,-m}\rangle=
\langle\Phi_{j_1;m,m}\Phi_{j_2;-m,-m}\rangle=\cr
&\pi\delta(j_1-j_2)C(j_1;k)
{\Gamma(1-{2j+1\over k})\Gamma(1+j+m)\Gamma(j-m+1)\Gamma(-2j-1)
\over
\Gamma({2j+1\over k})\Gamma(-j-m)\Gamma(m-j)\Gamma(2j+2)}~.\cr}}

\subsec{Two point functions in double scaled LST}

To use the results of the previous subsection for calculating
two point functions in double scaled little string theory we
have to construct BRST invariant observables in string
theory on
\eqn\fullst{{SL(2)_k\over U(1)}\times \IR^{d-1,1}\times LG(W=F)~.}
The analysis is very similar to that of \refs{\gkp,\ks}. We will
focus on (NS,NS) sector fields, since the rest of the observables
can be obtained from these by applying the spacetime supercharges.

The lowest lying states are the ``tachyons,''
described by the $(-1,-1)$ picture vertex operators
\eqn\tachver{T(k_\mu)=e^{-\varphi-\bar\varphi}e^{ik_\mu x^\mu}V_{j;m,m}~,}
where $k_\mu$ ($\mu=0,1,\cdots, d-1$) is the momentum along $\IR^{d-1,1}$
and $\varphi$, $\bar\varphi$ are the left and right moving bosonized
superghost fields. The mass-shell condition and GSO projection lead to
the following physical state constraints:
\eqn\physstcons{\eqalign{
&{1\over2}k_\mu k^\mu+{m^2-j(j+1)\over k}={1\over2}~,\cr
&m\in {k\over2}(2Z+1)~.\cr
}}
Another set of observables corresponds to gravitons, whose
$(-1,-1)$ picture vertex operators have the form
\eqn\gravver{e^{-\varphi-\bar\varphi}\xi_{\mu\nu}\psi^\mu\bar\psi^\nu
e^{ik_\mu x^\mu}V_{j;m,m}~,}
where $\psi^\mu$ are the worldsheet fermions on $\IR^{d-1,1}$,
$\xi_{\mu\nu}$ is the polarization tensor, and the physical state
constraints are
\eqn\physstgrav{\eqalign{
&{1\over2}k_\mu k^\mu+{m^2-j(j+1)\over k}=0~,\cr
&m\in kZ~.\cr}}
There are also transversality conditions on $\xi_{\mu\nu}$
which we will not specify.

The most general (NS,NS) sector observable is a linear combination
of vertex operators of the following form:
\eqn\mostgenobs{e^{-\varphi-\bar\varphi}
P_{N_L,N_R}(x^\mu,\psi^\mu,\bar\psi^\mu)
\tilde P_{\tilde N_L,\tilde N_R}(\psi^A,\bar\psi^A,J^A,\bar J^A)
V_{LG} e^{ik_\mu x^\mu}V_{j;m,m}~,}
where $P_{N_L,N_R}$ is a polynomail in
$\partial x^\mu,\partial^2 x^\mu,\cdots$,
$\bar\partial x^\mu,\bar\partial^2 x^\mu,\cdots$, $\psi^\mu$, $\bar\psi^\mu$
and their derivatives; $N_L$ and $N_R$ are its total
left and right-moving scaling dimensions, respectively. Similarly,
$\tilde P_{\tilde N_L,\tilde N_R}$ is a polynomial in
$J^A$, $\bar J^A$, their worldsheet
superpartners $\psi^A$, $\bar\psi^A$ and their derivatives,
with total scaling dimensions $(\tilde N_L,\tilde N_R)$.
$V_{LG}$ is a vertex operator from the $LG(W=F)$ sector
of \fullst. BRST invariance and the GSO projection require
\mostgenobs\ to be a bottom component of a worldsheet $N=1$ superfield,
and in addition to satisfy
\eqn\physstgen{\eqalign{
&{1\over2}k_\mu k^\mu+{m^2-j(j+1)\over k}+N_L+\tilde N_L
+\Delta_{LG}={1\over2}~,\cr
&Q_{LG}+{2m\over k}+F_L\in 2Z+1~,\cr}}
where $\Delta_{LG}$ and $Q_{LG}$ are the left-moving scaling dimension
and $U(1)_R$ charge of $V_{LG}$, respectively, and $F_L$ is the
left-moving fermion number of \mostgenobs\ (similar relations hold for
the right-movers). Moreover, the operator \mostgenobs\ must be invariant 
under the symmetry \Jga. Note also that $2m/k$ is not in general integer,
unlike \mbarm\ with $m=\bar m$. The reason for that is that the GSO 
projection acts as an orbifold on $SL(2)/U(1)$ (see \gkp), coupling 
it to the rest of the background \fullst.

We are now ready to compute correlation functions of the observables
\tachver, \gravver, \mostgenobs. Recall that correlation functions of
non-normalizable on-shell vertex operators in the bulk theory are interpreted
in the boundary theory as off-shell Green's functions \refs{\adscft,\abks}.
Thus, their analytic structure provides information about the spectrum
and interactions of the boundary theory.
The analytic structure of the two point function, for example,
contains information about the spectrum of the boundary theory.
We will next study it in our case.

For the general vertex operator \mostgenobs, the only non-trivial
part of the two point function is the worldsheet correlation
function $\langle V_{j;m,m}V_{j;-m,-m}\rangle$. This is the only
source of singularities of the amplitude, and thus we will focus
on it. As a first example, consider the two point function of the
tachyon field \tachver,
\eqn\twotach{\langle T(k_\mu)T(-k_\mu)\rangle=\pi C(j;k)
{\Gamma(1-{2j+1\over k})\Gamma(1+j+m)\Gamma(j-m+1)\Gamma(-2j-1)
\over
\Gamma({2j+1\over k})\Gamma(-j-m)\Gamma(m-j)\Gamma(2j+2)}~,}
where $j$, $m$ satisfy the constraints \posdim, \imprbound,
\physstcons. Note that the $\delta(j-j)$ from \twoptcoset\
has disappeared; it cancelled against a zero mode of the ghosts
which normally would make the two point function vanish in
string theory.

The two point function \twotach\ has a series of poles, which
we interpret as contributions of on-shell states in LST which
are created from the vacuum by the operator $T(k_\mu)$. We next
analyze these poles.

One series of poles corresponds to
\eqn\poleone{m=j+n;\;\;\;n=1,2,3,\cdots}
It is easy to check that these are single poles.
The residues of the poles all have the same sign,
in agreement with the expected unitarity of LST.
The corresponding masses of excitations are determined
by the first line in \physstcons:
\eqn\massbound{M_n^2=-k^\mu k_\mu=-1+{2\over k}
\left(j(2n-1)+n^2\right)~.}
For the smallest posssible value of $m$, $m=k/2$
(recall the second line in \physstcons\ and \poleone),
we find (using \imprbound) states with $n=1,2,\cdots, [k/2]$,
with masses
\eqn\mssn{M_n^2={2\over k}(n-1)(k-n)~.}
The lowest lying state is massless; it is followed by a
finite series of excited states. More generally, we have
\eqn\lbound{\eqalign{
&m={k\over2}l,\;\;\;l=1,3,5,7,\cdots\cr
&M_{n,l}^2=l-1+{2\over k}(n-1)(kl-n)~,\cr
&kl+1>2n>k(l-1)+1~.\cr
}}
The range of $n$ is obtained from \imprbound. Note that
Lorentz invariance of the LST implies that all the states
that can be created from the vacuum by the tachyon field
$T(k_\mu)$ (and whose masses are given by \lbound) have
spin zero.

The poles of \twotach\ which are
related to \poleone\ by $m\to-m$ are not independent
of those discussed above. All other poles violate
one or more of the bounds\foot{Note,
for example, that if we imposed the constraint \unitbound\ rather
than \imprbound, we would be lead to the rather unpleasant
prediction that LST has tachyons. The pole of \twotach\ at $m=k/2$,
$j=(k-1)/2$ leads \physstcons\ to $M^2=-1+{1\over 2k}$, which
is in general negative. Furthermore, if $k$ is integer \twotach\
has a {\it double} pole at that value of $j$, which seems difficult
to interpret in LST.} \posdim, \imprbound.

The massless state corresponding to $n=l=1$ in \lbound\ has a
natural interpretation in the fivebrane theory. We will describe
it in the six dimensional maximally supersymmetric case, but 
the discussion can probably be extended to general singularities
\noncomp. As we saw in section 1, six dimensional double scaled 
type IIB LST describes the dynamics of $N=k$ fivebranes at the point
\pointmod\ in the Coulomb branch. Due to the Higgs mechanism, the
only massless states one expects to find in this theory are $N-1$
supermultiplets corresponding to the Cartan subalgebra of $SU(N)$. 
The massless state with $n=1$ in \mssn\ belongs to one of these $N-1$
supermultiplets\foot{It is sufficient to exhibit any one of the members
of the supermultiplet, since the supersymmetry structure, worked out
in \gkp, then guarantees the appearance of the rest of the supermultiplet,
with the right quantum numbers.}.

It is not difficult to exhibit the other $N-2$ massless multiplets.
For this, we need to examine the two point functions of tachyons
\tachver\ which have a non-trivial wave function in the minimal model
denoted by $LG(W=F)$ in \fullst. In our case the superpotential is \anmo
\eqn\supsix{W=z_1^N+z_2^2+z_3^2~,}
corresponding to an $N=2$ minimal model. The chiral worldsheet
operators can be written as $z_1^i$, with $i=0,1,\cdots, N-2$. 
A natural generalization of the tachyon vertex \tachver\ is
\eqn\tikmu{T_i(k_\mu)=e^{-\varphi-\bar\varphi}e^{ik_\mu x^\mu}z_1^i
V_{j;m,m}~,}
where
\eqn\physzz{\eqalign{
&{1\over2}k_\mu k^\mu+{m^2-j(j+1)\over k}+{i\over 2N}={1\over2}~,\cr
&m={k\over2}l;\;\; l\in -{i\over N}+2Z+1~,\cr
}}
which is simply \physstgen\ for this particular case. The tachyon
$T(k_\mu)$ \tachver\ is  in fact $T_0(k_\mu)$.
The two point function of $T_i$ is proportional to that of
$T_0$, \twotach, and has poles giving rise to a spectrum
similar to \lbound. It is easy to check that there is a physical
pole at $m=kl/2$, $l=1-{i\over N}$, $m=j+1$, which gives rise
to a massless state in six dimensional LST. There are thus precisely
$N-1$ massless states, corresponding to $i=0,\cdots, N-2$, in agreement
with the above expectations from $SU(N)$ gauge theory.

Moving on to the two point function of the graviton, we see that
in this case $m=kl/2$ with $l\in 2Z$. The two point function again
has poles for $j$ satisfying \poleone, which using \physstgrav\
corresponds to the mass spectrum
\eqn\massgrav{\eqalign{
&M_{n,l}^2=l+{2\over k}(n-1)(kl-n)~,\cr
&l=2,4,6,\cdots\cr
&kl+1>2n>k(l-1)+1~.\cr}}
All the states that the bulk graviton \gravver\ couples to are
massive\foot{The two point function of the graviton provides an example of the
importance of the constraint \posdim. Ignoring it, one finds a series
of poles of \twoptcoset\ at $m=0$, $j=(n-1)/2$, $n=2,4,6,\cdots$,
which, if present, would correspond to tachyons in LST with masses which
can be read off \physstgrav, $M^2=-2j(j+1)/k$.}. 
This is consistent with the fact that the boundary
theory is not a theory of gravity -- the spectrum does not
contain a massless spin two particle.

Consider, finally, the two point function of the most general
on-shell vertex operator \mostgenobs. Let $N\equiv N_L+\tilde
N_L+\Delta_{LG}$ be the total excitation level corresponding
to this vertex operator. As before, the poles correspond to
\poleone, which in this case leads to the spectrum
\eqn\massgen{\eqalign{
&M_{n,l}^2=2N+l-1+{2\over k}(n-1)(kl-n)~,\cr
&kl+1>2n>k(l-1)+1~,\cr}}
where $l$ is defined as in the first line of \lbound,
$m=kl/2$. It runs over the range implied by the second line
in \physstgen, $l\in -Q_{LG}-F_L+2Z+1$, subject to the unitarity
constraint
\eqn\unitconsst{{m^2-j(j+1)\over k}+\Delta_{LG}>0}
which generalizes \posdim\ to operators which have a non-trivial
projection in $LG(W=F)$.

\subsec{Three point functions}

Three point couplings between the states \massgen\ can be obtained
from the three point function in the usual way (by taking the
external legs on-shell and computing the residue of the resulting
poles).
To find the three point functions in the double scaled LST
one follows the same steps as in the computation of the two point
functions. We will only present the procedure schematically, leaving
the details for future work.
First, one needs the three point functions of the $SL(2)$ primaries
\ooyy\ in the $H_3^+$ CFT. Those are given in \teschner:
\eqn\thpf{\langle \Phi_{j_1}(x_1,\bar x_1)\Phi_{j_2}(x_2,\bar x_2)
\Phi_{j_3}(x_3,\bar x_3)\rangle
=D(j_1,j_2,j_3;k)C(j_1,j_2,j_3;x_1,x_2,x_3)~,}
where
\eqn\cjjj{C(j_1,j_2,j_3;x_1,x_2,x_3)=|x_1-x_2|^{2(j_3-j_1-j_2-1)}
|x_1-x_3|^{2(j_2-j_1-j_3-1)}|x_2-x_3|^{2(j_1-j_2-j_3-1)}~.}
$D(j_1,j_2,j_3;k)$ does not depend on $x$, and its explicit
form is given in \teschner. To compute the three point
functions in the coset theory we need to transform the
fields $\Phi_{j_i}(x_i,\bar x_i)$, $i=1,2,3$, to the modes 
\modeops; this leads to integrals of hypergeometric functions 
depending on $j_i,m_i$.
Finally, the three point functions in the spacetime theory are
obtained by following the same steps as in subsection 3.2.

\newsec{Discussion}

By defining Little String Theory in a double scaling limit (see
section 2), we were able to study the theory at weak coupling, $1/x$
\scvar. We used this description to compute correlation
functions in the theory to leading order in $1/x$.

One of the main results of this paper is the calculation
of the two point functions of observables in double scaled
LST (\twotach\ and its generalization to other vertex
operators), and the corresponding spectrum of masses \massgen.

The two point functions we computed have a relatively simple
analytic structure. They exhibit a
series of single poles which we interpreted as arising from
single particle states in LST, created from the vacuum by
applying different operators\foot{As we pointed out, a nice
consistency check is that the residues of all poles in the
two point functions have the same sign, which is
necessary for unitarity of LST.}. This might come as a surprise,
since in general one would expect two point functions in a
non-trivial theory such as LST to be complicated functions
of the momenta. The eigenstates of the Hamiltonian need not
be interpretable in terms of free particles, and one would
expect a complicated structure containing resonances, thresholds
(branch cuts) and ``brown muck.''

Of course, crucial to the simplicity of the two point functions
is the fact that the theory is weakly coupled. As mentioned
above, we only computed the two point function to leading order
in $1/x$, and it is reasonable to expect the theory to be free
in this limit. A good analogy is confining $SU(N)$ gauge theory
in four dimensions. For finite $N$ the spectrum of the theory
is complicated and, correspondingly, the two point functions
of gauge invariant observables are expected to have a complicated
analytic structure. As $N\to\infty$ the theory simplifies and becomes
essentially a free field theory of the bound states (glueballs).
The two point functions of observables like ${\rm Tr} F_{\mu\nu}F^{\mu\nu}$
are expected to simplify in the limit and have an analytic structure
similar to the one found here.

Another interesting fact about the two point functions is that
while the full spectrum of masses we find is rather rich,
most operators do not couple to most of the states! There are two
aspects to this, one that seems natural, and a second one that
looks more surprising.

The natural one is the following. Even
if the spectrum of LST at large $x$ consists of free particles
with the masses \massgen, one would in general expect that in the
absence of special symmetries, generic observables can create from
the vacuum both single and multi-particle states. The latter would
contribute branch cuts in the two point functions. The fact that we
only find poles means that the observables that we study, on-shell
vertex operators in the bulk theory, only couple to single particle
states in LST. This seems natural, since they are single string
states in the bulk, and we are performing an expansion in $g_s$. In the
gauge theory analogy mentioned above, fundamental
string vertex operators are analogs of single trace operators in
large $N$ Yang-Mills theory.

More surprisingly, we find that even within the set of single
particle operators/states, most operators do not couple to
most states. The tachyon operators $T(k_\mu)$
\tachver\ only couple to states \massgen\ with $N=0$, the
gravitons \gravver\ couple to states with $N=1/2$, etc. This
is not expected to happen in general. E.g. in the large $N$ 
gauge theory example the spectrum is believed to exhibit
Hagedorn growth, and one expects that all single trace
operators have a non-zero overlap with all states which
have the same quantum numbers (\eg\ the same spin). For
example, the operator ${\rm Tr} F_{\mu\nu}F^{\mu\nu}$
mentioned above, should be able to create from the vacuum
all states with spin zero.

The fact that in LST we find an infinite number of additional
selection rules suggests that the weakly coupled theory has
a very large symmetry group. It would be nice to understand
it better.

It is natural to suggest that in LST there is a
state-operator correspondence, namely, that each operator
in the bulk can create a particular one particle state in
the boundary theory. For example, a tachyon vertex operator
\tachver\ with given $n$, $l$ (or $j$, $m$) can create
from the vacuum a unique state, with mass given by \lbound.
If this proposal is correct, it is easy to compute the
degeneracy of states with any mass \massgen. Since the mass
spectrum \massgen\ is very similar to that of the bulk theory,
and the state-operator correspondence suggests that the
degeneracies are the same as well, one finds a Hagedorn density
of states
\eqn\rhoee{\rho(E)\sim e^{\beta_HE}~,}
where $\beta_H$, the inverse Hagedorn temperature, is the same as
that of the bulk theory. We emphasize that this last conclusion
assumes the state-operator correspondence, and should be examined
more carefully.

Note that all the poles of the two point functions that give rise
to the masses \massgen\ occur for values of $(j,m)$ that satisfy
\poleone\ (or, equivalently, its mirror image related by $m\to -m$).
Algebraically, these are values for which the underlying $SL(2)$
representations are one sided. They contain a state, $m=j+1$, which
is annihilated by $J^-$ (or $m=-(j+1)$, annihilated by $J^+$). These
representations are sometimes refered to as the principal discrete
series. A natural question is why does the principal discrete
series play a special role in the analysis, and how one can think
of the resulting spectrum of the LST.

\lref\gks{A. Giveon, D. Kutasov and N. Seiberg,
Adv. Theor. Math. Phys. {\bf 2} (1998) 733, hep-th/9806194.}%
\lref\mssm{N. Seiberg, Prog. Theor. Phys. Supp. {\bf 102} (1990) 319.}

We would like to make two comments regarding these issues. The first
is that the appearance of the principal discrete series is not
particularly surprising if we compare to the better understood
case of string theory on $AdS_3$ \refs{\adscft,\gks}. In that
case, the holographically dual spacetime theory is a two dimensional CFT
and the operators $\Phi_j(x,\bar x)$ \ooyy\ correspond to local
operators in this theory.

While there are well known subtleties \refs{\mssm,\teschner}
with the state-operator correspondence in CFT on $AdS_3$,
the {\it spacetime} CFT is a more or less conventional unitary
CFT, and in particular it does have a state-operator correspondence.
The state corresponding to a given operator $V\Phi_j(x,\bar x)$
($V$ is a vertex operator which depends on the rest of the string
background; see \gks\ for the details) is obtained by acting with
$V\Phi_j(x,\bar x)$ on the vacuum. In radial quantization the `in' vacuum
corresponds to $x=0$, hence incoming states correspond to $\lim_{x\to 0}
V\Phi_j(x,\bar x)|0\rangle$.

Expanding $\Phi_j(x,\bar x)$ in modes, as in \modeops, we see that
the mode that creates the above state in the spacetime CFT has
$m=\bar m=-j-1$. Modes with larger $m$ must kill the vacuum, while those
with smaller $m$ do not contribute in the limit $x\to 0$. Similarly,
there are states created by acting on the vacuum with
$\partial_x^{n-1}\partial_{\bar x}^{n-1}\Phi_j(x,\bar x)$.
The modes that create these states have $m=-j-n$. The modes
\poleone\ discussed in section 3 are related to them by $m\to -m$;
they create states when acting on the `out' vacuum at $x\to\infty$.

Thus, at least in string theory on $AdS_3$ the modes \poleone\
corresponding to the principal discrete series certainly do
create physical states in the spacetime theory. It is perhaps
not surprising that something very similar seems to be
going on in the $SL(2)/U(1)$ case.

\lref\dvv{R. Dijkgraaf, E. Verlinde and H. Verlinde, Nucl. Phys.
{\bf B371} (1992) 269.}%

The second comment concerns the physical interpretation of the spectrum
\massgen. It is known \dvv\ that the dynamics of winding modes
on the cigar $SL(2)/U(1)$ is equivalent by T-duality to the motion
of a particle in a potential which is attractive for small $\phi$ and
goes to a constant at large $\phi$.
The spectrum contains bound states corresponding to the principal
discrete series, and a continuum above a gap, which corresponds
to the normalizable states with $j=-{1\over2}+i\lambda$. The discrete
states can presumably be thought of in the original variables as some
sort of bound states whose wave function is supported near the tip of
the cigar. The states with $j=-{1\over2}+i\lambda$ make wave packets
that can live arbitrarily far from the tip of the cigar and have an
arbitrary momentum $\lambda$ (hence the continuum). It is easy
to verify using \physstgen, \massgen\ that the continuum starts right
above the highest mass discrete state (in a sector with given $N$, $l$),
as expected from the form of the potential for the winding modes \dvv.

One can ask whether the normalizable states with $j=-{1\over2}+i\lambda$
should be included in the LST. For this to be the case, it seems to us
necessary that there are {\it some} observables in the theory which
can create these states from the vacuum. The class of observables
analyzed in this paper, the non-normalizable fundamental string operators,
does not seem to couple to these states. If no other observables do
either, we would be inclined to believe that these modes decouple
from the LST dynamics\foot{This issue is further confused by the
fact \refs{\teschner,\mssm} that in correlation functions of non-normalizable 
operators in CFT on $SL(2)/U(1)$ one is typically instructed to sum over
all the normalizable states in internal channels. Nevertheless, it seems
that the singularity structure of the resulting correlation functions
can be interpreted purely in terms of states bound to the tip of the cigar.}.

\lref\ahbnks{J. Maldacena and A. Strominger, hep-th/9710014,
JHEP {\bf 9712} (1997) 008;
O. Aharony and T. Banks, hep-th/9812237, JHEP {\bf 9903}
(1999) 016.}%

Returning to the spectrum \rhoee, our conclusion that the
density of states of the LST grows exponentially with energy,
with the same Hagedorn temperature as that of the bulk theory,
seems to be at odds with the discussion of \ahbnks, 
where it was argued that the Hagedorn temperature is in fact lower 
by (roughly) a factor of $\sqrt k$ (and thus the density of states
$\rho$ grows more rapidly with energy than \rhoee). 

\lref\rr{A. Rajaraman and M. Rozali, hep-th/9909017.}%

Our attitude to this apparent discrepancy is the following.
The analysis of \ahbnks\ applies in our language to energies
$E>>m_W\sim x$, while we computed the spectrum of excitations
with energies $E<<m_W$. We find a Hagedorn growth of the density
of states \rhoee\ with a smaller $\beta_H$ than that of \ahbnks\
in the range $m_s<<E<<m_W$. It is reasonable to expect that the
density of states with $E>>m_W$ is much higher and agrees with
\ahbnks. 

The regime $E>>m_W$ is strongly coupled in our description and is
difficult to study directly. 
As has been learned in many examples
in the last few years, typically only states that are protected by
supersymmetry can be reliably counted at strong coupling. Perhaps
D-brane states 
which preserve some of the supersymmetry
in the background \fullst\ would resolve the discrepancy
(see \rr\ for a recent discussion).

Finally, some of the vacua of LST
that our discussion applies to correspond to theories on 
fivebranes wrapped around Riemann surfaces, which are relevant
for describing strongly coupled gauge theories using branes \gkp.
We hope that our results will help improve this understanding.

\bigskip
\noindent{\bf Acknowledgements:}
We thank O. Aharony, T. Banks, O. Pelc, M. Porrati, J. Teschner, C. Vafa and
A. B. Zamolodchikov for useful
discussions. We also thank the ITP in Santa Barbara for hospitality during
the course of this work. D.K. thanks the HET group at Rutgers
University for hospitality. This research was supported in part by NSF grant
\#PHY94-07194. The work of A.G. is supported in part by the Israel Academy
of Sciences and Humanities -- Centers of Excellence Program, and by
BSF -- American-Israel Bi-National Science Foundation. D.K. is supported
in part by DOE grant \#DE-FG02-90ER40560.

\listrefs
\end